\documentclass[preprint,aps]{revtex4}
\usepackage{graphicx}
\usepackage{epstopdf}
\usepackage{dcolumn}
\usepackage{bm}
\usepackage{color}

\begin{document}


\title{Graphene on metallic surfaces: problems and perspectives}

\author{E. N. Voloshina$^{1,}$\footnote{Corresponding author. E-mail: elena.voloshina@fu-berlin.de} and Yu. S. Dedkov$^{2}$}

\affiliation{$^1$Physikalische und Theoretische Chemie, Freie Universit\"at Berlin, 14195 Berlin, Germany}
\affiliation{$^2$SPECS Surface Nano Analysis GmbH, Voltastra\ss e 5, 13355 Berlin, Germany}

\date{\today}

\begin{abstract}

The present manuscript summarizes the modern view on the problem of the graphene-metal interaction. Presently, the close-packed surfaces of $d$ metals are used as templates for the preparation of highly-ordered graphene layers. Different classifications can be introduced for these systems: graphene on lattice-matched and graphene on lattice-mismatched surfaces where the interaction with the metallic substrate can be either ``strong'' or ``weak''. Here these classifications, with the focus on the specific features in the electronic structure in all cases, are considered on the basis of large amounts of experimental and theoretical data, summarized and discussed. The perspectives of the graphene-metal interface in fundamental and applied physics and chemistry are pointed out.
\end{abstract}

\maketitle

\section{Introduction}

\begin{figure}[t]
\centering
\includegraphics[width=0.6\textwidth]{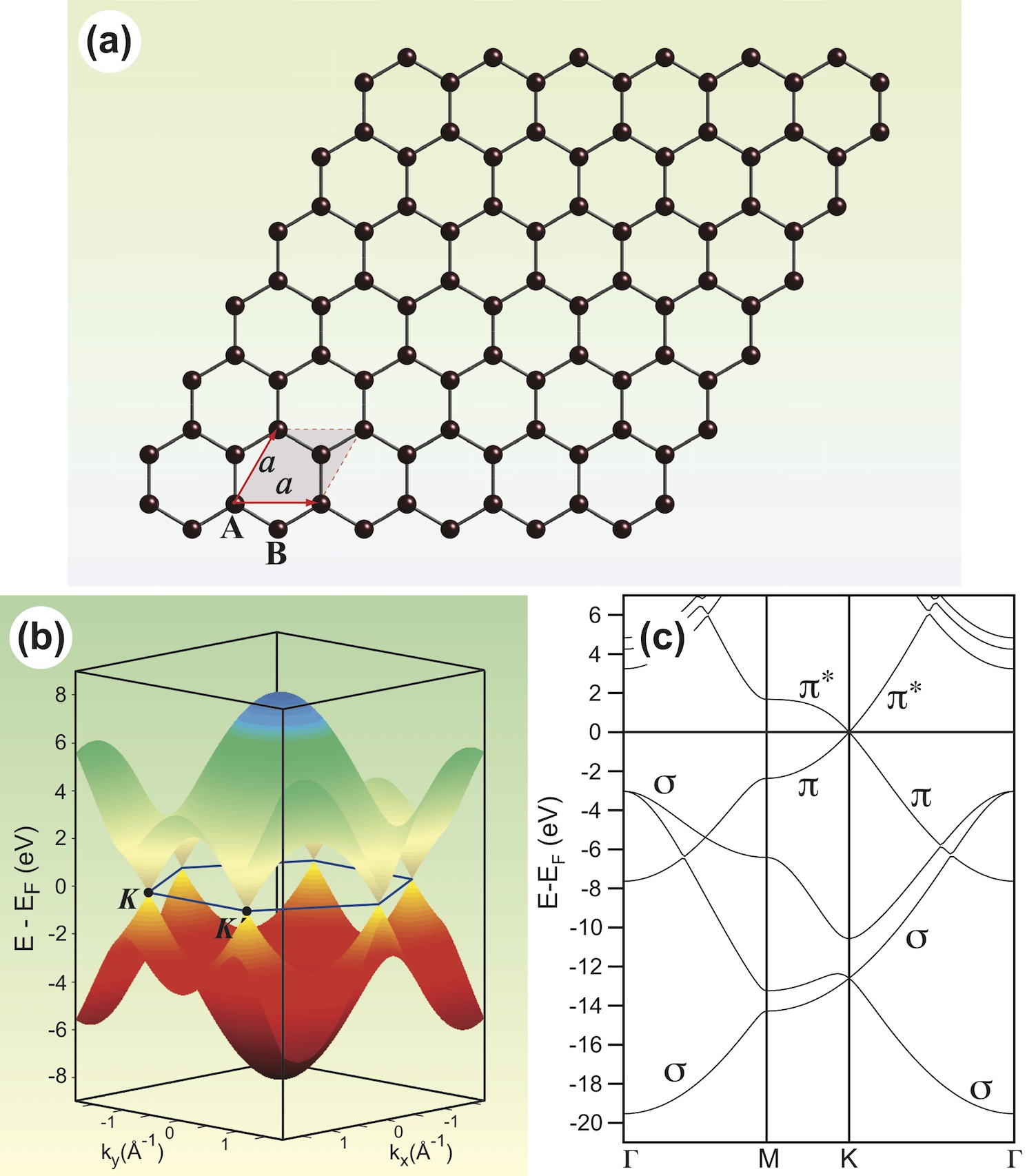}\\
\caption{\label{fig:graphene} (a) Crystal structure of the graphene layer where carbon atoms are arranged in a honeycomb lattice. The unit cell of graphene with lattice constant $a$ has two carbon atoms per unit cell, A and B. (b) Electronic dispersion of $\pi$ and $\pi^*$ states in the honeycomb lattice of free-standing graphene obtained in the framework of tight-binding approach. These branches have linear dispersion in the vicinity of the $K$ points of the Brillouin zone of graphene. (c) Band structure of free-standing graphene as obtained by means of DFT ($\sigma$, $\pi$, and $\pi^*$ bands are marked).}
\end{figure}

The physics and chemistry of graphene overlayers on metallic surfaces have become the focus of worldwide research since the extraordinary properties of the single layer graphene (Fig.~\ref{fig:graphene}) were demonstrated in 2004~\cite{Novoselov:2004a,Novoselov:2005,Zhang:2005}. This topic is interesting from several points of view. From the practical side, as was recently demonstrated, the metallic substrate can be used for the preparation of graphene layers of different thicknesses with the extraordinary quality that can be transferred onto an insulating or polymer support~\cite{Yu:2008,Kim:2009a,Li:2009,Bae:2010}. In the latter case the obtained material was successfully used for the fabrication of flexible touch screens~\cite{Bae:2010}. These facts together with the relative easiness of the preparation procedure of graphene on metals and its transfer onto the polymer or insulating support make this method a most promising one for the graphene industry. Additionally, in any of the technical applications the metallic contacts to a graphene layer determine the doping of graphene (and, as a consequence, its transport properties) that points the importance of studies of the graphene/metal interface~\cite{Lee:2008,Khomyakov:2009,Khomyakov:2010}. 

From the other, more fundamental, point of view, the nature of bonding itself at the graphene-metal interface is still far from a complete understanding. For example, the weakening of the bonding between graphene and $d$-metals from the 4th and 5th periods of the periodic table can be related to the filling of the corresponding $d$ shell. This leads to a decrease of the degree of hybridization between $d$ states of the metal and the  $\pi$ states of graphene (effect of hybridization has to fulfill two conditions simultaneously: spatial and energetic overlapping of orbitals). The representative example here is the difference of the graphene behavior on the Ni(111) and Cu(111) surfaces. Both interfaces are lattice-matched. The graphene/Ni(111) interface is the representative 
example of the so-called ``strong'' interaction between graphene and the metallic substrate with the significant hybridization of the graphene\,$\pi$ and Ni\,$3d$ states that leads to the complete destruction of the graphene Dirac cone~\cite{Bertoni:2004,Khomyakov:2009,Khomyakov:2010,Weser:2011}. In the case of ``weakly'' bonded graphene/Cu(111) the theoretical description of this system [the small lattice-mismatch of $3.9$\% between graphene and Cu(111) is omitted] shows that the graphene-derived Dirac cone is preserved in the electronic structure~\cite{Khomyakov:2009} and its position is defined by the doping level. These observations are explained by the absence of hybridization around the Fermi level ($E_F$) between Cu\,$3d$ and graphene\,$\pi$ states. A similar approach can be used for the description of bonding mechanisms of graphene on surfaces of $4d$ and $5d$ transition metals. 

The situation becomes more intriguing if one now considers the interaction of graphene with the $d$-metals from the same group. Good examples are the Co-Rh-Ir or Ni-Pd-Pt sequences. In both of them the strength of interaction of graphene with a metallic surface is decreased. Usually, as a measure of the strength of the graphene interaction with metallic surfaces one uses the bonding energy per carbon atom, the state of the Dirac cone or the doping level which is defined by the position of the Fermi level with respect to the Dirac point. It is interesting to note that in all graphene-metal systems the bonding energy of graphene is too small compared to the value which can be estimated from the temperature needed for the graphene synthesis, e.\,g. by chemical-vapor deposition (CVD) method. Here the synthesis temperature is varied from $450^\circ$C for Co(0001) to $1200^\circ$C for Ir(111), that is opposite to the current description~\cite{Wintterlin:2009,Batzill:2012,Dedkov:2012book}. Following the simple estimation for the energy $E=k_BT$, graphene has to be bonded stronger to Ir(111) compared to Co(0001), that is opposite to results obtained from the theoretical calculations: the graphene layer on Co(0001) is considered as a ``strongly'' interacting system with a bonding energy of $0.132$\,eV/C-atom, a high doping level of graphene, and a fully destroyed Dirac cone due to the strong hybridization of Co\,$3d$ and graphene $\pi$ states~\cite{Eom:2009}. Graphene on Ir(111) is a ``weakly'' bonded system with a bonding energy of $0.05$\,eV/C-atom, very low doping level of graphene, and an intact Dirac cone~\cite{Busse:2011}. From the other side, taking into account the higher catalytic activity of the Co surface compared to Ir one we can expect that graphene will grow faster on Co(0001) at higher temperatures. However, this is not the case as was demonstrated in Ref.~\cite{Gruneis:2009} where the decreasing of the graphene growth rate was observed at temperatures higher than $800^\circ$\,C. Most probably this is due to the instability of graphene on Ni(111) and Co(0001) at such high temperatures. At the same time the mismatch between a graphene layer and the metallic substrate can not be considered as an argument explaining the bonding of graphene on metals: Rh(111) and Ir(111) substrates both do not match to the graphene plane. However, graphene/Rh(111) is a ``strongly'' bonded system opposite to graphene/Ir(111)~\cite{Wang:2010ky,Voloshina:2012a}.

\begin{figure*}
\centering
\includegraphics[width=1\textwidth]{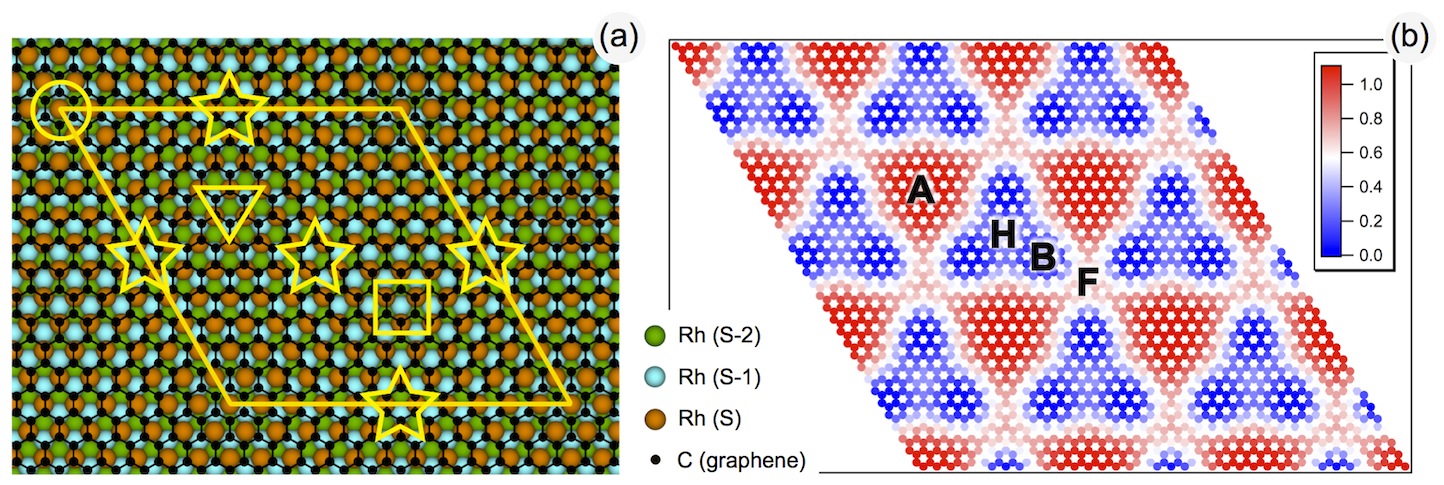}\\
\caption{\label{fig:grRh_struct} (a) Crystallographic structure of the graphene/Rh(111) interface. The high-symmetry stacking positions of carbon atoms on Rh(111) are marked by different symbols (see text for details). (b) Height variation of carbon atoms in the graphene/Rh(111) system. The zero level is taken for the $BRIDGE$-position.}
\end{figure*}

Another interesting question related to the graphene-metal interface is the existense or absence of the energy gap in the energy dispersion for the graphene $\pi$ states around the Dirac point. Generally, the appearance of such an energy gap is connected with the breaking of the sublattice symmetry in the graphene layer. However the carbon atoms of the graphene unit cell occupy nonequivalent positions in all cases of the graphene adsorption on a metallic substrate. Thus, one should expect violation of the sublattice symmetry and, as a consequence, the appearance of an energy gap in the electronic spectrum of graphene. In fact, the situation is not so unequivocal. Whereas, for the ``strongly'' bonded graphene on metals the appearance of such a gap is more or less obvious (due to the strong hybridization between valence band electronic states of graphene and metal and high level of doping) and was observed experimentally (and confirmed theoretically) for graphene on Ni(111), Co(0001), or Ru(0001)~\cite{Gruneis:2008,Dedkov:2010a,Brugger:2009}, for the ``weakly'' bonded graphene the experimental data are inconsistent. For example, some experimental results for graphene on Au(111) or Cu(111) indicates the existence of such a gap~\cite{Enderlein:2010,Varykhalov:2010a}, whereas for the graphene/Al(111) system experimental data demonstrate no gap~\cite{Giovannetti:2008,Khomyakov:2009}. The theoretical calculations predict the gap-less situation in all cases of the ``weakly'' bonded graphene on metals~\cite{Giovannetti:2008,Khomyakov:2009}.

The present article is intended to give a timely account of recent developments in the investigation of the graphene-metal systems, in particular the aspects of the interaction at the interface will be highlighted. These points will be reviewed taking into account the massive set of theoretical and experimental data. The paper is organized as follows. The introductory part contains the main questions which arise in the course of the study of the graphene-metal interface. The modern preparation methods as well as computational and experimental approaches are briefly reviewed in the second section. Sections\,III and IV demonstrate several representative examples of the graphene-metal systems which are used for the critical discussion of the existing problems rising in the introductory part and others. Conclusions are given in Section\,V.

\section{Experimental and theoretical approaches}

Many references on the particular experimental and theoretical works as well as on the preparation methods of graphene on metals can be found in recent review articles~\cite{Wintterlin:2009,Batzill:2012,Dedkov:2012book}. Here we briefly overview the main experimental and theoretical methods used for the investigation of the graphene/metal systems. 

Presently there are two commonly used ways of the preparation of graphene on metals~\cite{Yu:2008,Kim:2009a,Li:2009,Bae:2010}. Both of them include the reaction of the metallic surfaces with hydrocarbons. In the first case the metallic substrate is kept at high temperature and pressure of hydrocarbons during long time that leads to the dissolving of carbon atoms in the bulk metal. The cooling of the metallic substrate leads to the segregation of the carbon atoms at the surface of the metal. The careful control of temperature, pressure and the rate of cooling can produce high quality graphene, from one layer to multilayers. This method can be successfully used for the preparation of graphene layers of controllable thickness on polycrystalline metallic support (mainly on Ni, but was also demonstrated for Cu) and it was demonstrated that graphene obtained in such a way and transferred on the polymer insulating support can be used in the industrial applications for fabrication of gas sensors or touch screens. The second method uses the high catalytic activity of low-index metallic substrates and that the single graphene-layer growth is a self-limiting process due to the inertness of the graphene surface~\cite{Nagashima:1994a}. Here the surface of metal heated to the particular temperature is exposed to the hydrocarbon atoms at the pressure of $10^{-8}-10^{-5}$\,mbar allowing the formation of single graphene layers of very high quality. 

\begin{figure*}
\centering
\includegraphics[width=1\textwidth]{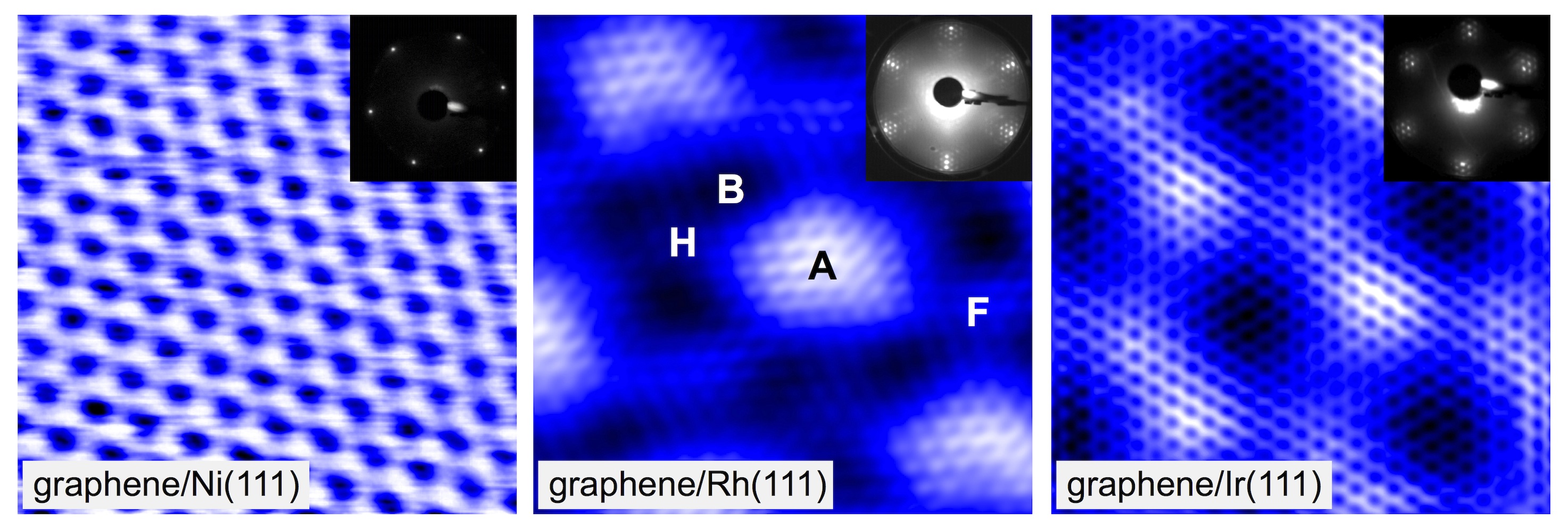}\\
\caption{\label{fig:stm} STM images of a graphene layer on Ni(111), Rh(111), and Ir(111). The corresponding LEED images are shown as insets. The capital letters in the middle STM image denote the high-symmetry adsorption sites of carbon atoms on Rh(111).}
\end{figure*}

The graphene/metal system is an ideal object for the surface science studies due to the two-dimensional geometry of the system and consequently the character of the electronic states. Initially the problem of the graphene layer (or monolayer of graphite in former time) on metal was connected with the catalytic properties of metallic surfaces and the passivation properties of graphene. It was found that adsorption or segregation of carbon layers leads to the quenching of the catalytic activity of metallic surfaces. Correspondingly, focusing on the applied research, most of the studies were performed with the application of low-energy electron diffraction (LEED) and Auger electron spectroscopy (AES) methods with rare applications of photoelectron spectroscopy. Several applications of (high-resolution) electron-energy loss spectroscopy methods to graphene/metal systems were published with the aim of determination of the strength of interaction between carbon layer and metallic surface. With the discovery of the fascinating properties of the free-standing graphene, the main focus of the surface science studies was moved to the investigation of crystallographic structure and electronic properties of the graphene/metal systems by means of scanning tunneling microscopy (STM) and angle-resolved photoelectron spectroscopy (ARPES), respectively. Here, due to the fact that in many cases of the prepared graphene/metal interface the dispersion of graphene-derived $\pi$ states is strongly disturbed, compared to free-standing graphene, a large number of works were devoted to the modification of such systems via adsorption or intercalation of different materials on top or underneath of the graphene layer. Here we would like to address the readers to the several recent reviews on graphene/metals systems where many experimental works are cited~\cite{Wintterlin:2009,Batzill:2012,Dedkov:2012book}. The corresponding references on particular works will be introduced later in the discussion.

The most widely used theoretical approach, when studying the graphene/metal systems, is density functional theory (DFT), ranging from the $X_\alpha$ method~\cite{Souzu:1995} to the recently revisited random phase approximation (RPA)~\cite{Mittendorfer:2011,Olsen:2011fw}.  Initially, the lattice-matched graphene/metal systems were in the focus of these studies~\cite{Bertoni:2004}. Parallel execution of DFT-codes, like VASP, available for many users allows to study the graphene/metal lattice mismatched systems, where graphene and metal layer form moir\'e structures with periodicity of ca.\,$3$\,nm and the model unit cells contain few hundreds of atoms~\cite{Wang:2008}. Still, most of the studies were performed employing the local density approximation (LDA), yielding reasonable binding energies (due to error cancellation) contrary to the standard generalized gradient approximation (GGA). In the latter case, the weakly bonded structures are usually predicted to be less stable than in reality, and sometimes they cannot even be localized as a stationary point on the potential energy surface~\cite{Mittendorfer:2011,Olsen:2011fw}. (It is necessary to note here, that density functional theory itself is capable of providing the exact solution to the Schr\"odinger equation, including long-range correlations - the dispersion. The root of the rather unsatisfactory description of the intermolecular interactions in DFT lies in the approximations made in the DFT functionals of all types: LDA, GGA, hybrid functionals, and meta-GGA functionals).

A new era in the computational chemistry of graphene/metal systems began with the development of approaches allowing to account for dispersion interactions, which are extremely important when investigating layered materials. A variety of methods were suggested to mitigate the inherent problem of locality in DFT (for a review, see Ref.~\cite{Riley:2010}). Here we will name only approaches which have been employed so far for the investigation of graphene/metal interfaces (see e.g. Refs.~\cite{Adamska:2012,Busse:2011}): (i) the dispersion correction schemes (referred to as DFT-D) by Grimme~\cite{Grimme:2006} and Tkatchenko and Scheffler~\cite{Tkatchenko:2009};  (ii) utilization of truly nonlocal functionals, e.g. the so-called vdW-DF~\cite{Dion:2004}.

\section{Examples of the graphene/metal interfaces}

\subsection{Graphene on Ni(111), Ir(111), and Rh(111)}

As an example of the graphene/metal system let us consider the representative case of this interface, namely graphene/Rh(111). In the unit cell of this system the $(12\times12)$ graphene layer is arranged on the $(11\times11)$ Rh(111) stack~\cite{Wang:2008,Wang:2010ky,Wang:2011hh,Voloshina:2012a}. This example we will use for the discussion of all features of the graphene-based interface with metal. Generally, this system is a good example of the lattice-mismatch interface between a graphene layer and the metallic surface. Its crystallographic structure is shown in Fig.~\ref{fig:grRh_struct}. Considering the arrangement of the graphene layer on Rh(111) one can identify several high-symmetry stacking positions for carbon atoms in the layer. (Different notations are used in the literature to mark these positions.) They are:
\begin{itemize}

\item $ATOP$-position -- carbon atoms surround the metal atom of the top layer and are placed in the $hcp$ and $fcc$ hollow positions of the Rh(111) stack above $(S-1)$ and $(S-2)$ Rh-layers, respectively ($hcp-fcc$ position) [circle in Fig.~\ref{fig:grRh_struct}(a)];

\item $FCC$-position -- carbon atoms surround the $fcc$ hollow site of the Rh(111) surface and are placed in the $top$ and $hcp$ hollow positions of the Rh(111) stack above $(S)$ and $(S-1)$ Rh-layers, respectively ($top-hcp$ position) [rectangle in Fig.~\ref{fig:grRh_struct}(a)];

\item $HCP$-position -- carbon atoms surround the $hcp$ hollow site of the Rh(111) surface and are placed in the $top$ and $fcc$ hollow positions of the Rh(111) stack above $(S)$ and $(S-2)$ Rh-layers, respectively ($top-fcc$ position) [triangle in Fig.~\ref{fig:grRh_struct}(a)];

\item $BRIDGE$-position -- carbon atoms are bridged by the Rh atom in $(S)$ layer [star in Fig.~\ref{fig:grRh_struct}(a)].

\end{itemize}
\noindent All these arrangements of carbon atoms on transition-metal surfaces can be easily identified in the experiment, for lattice-matched as well as lattice-mismatched systems. The representative examples are shown in Fig.~\ref{fig:stm}, where the STM images of a graphene layer on Ni(111), Rh(111), and Ir(111) are shown. The corresponding LEED images are presented as the insets.

\begin{figure}[t]
\centering
\includegraphics[width=0.6\textwidth]{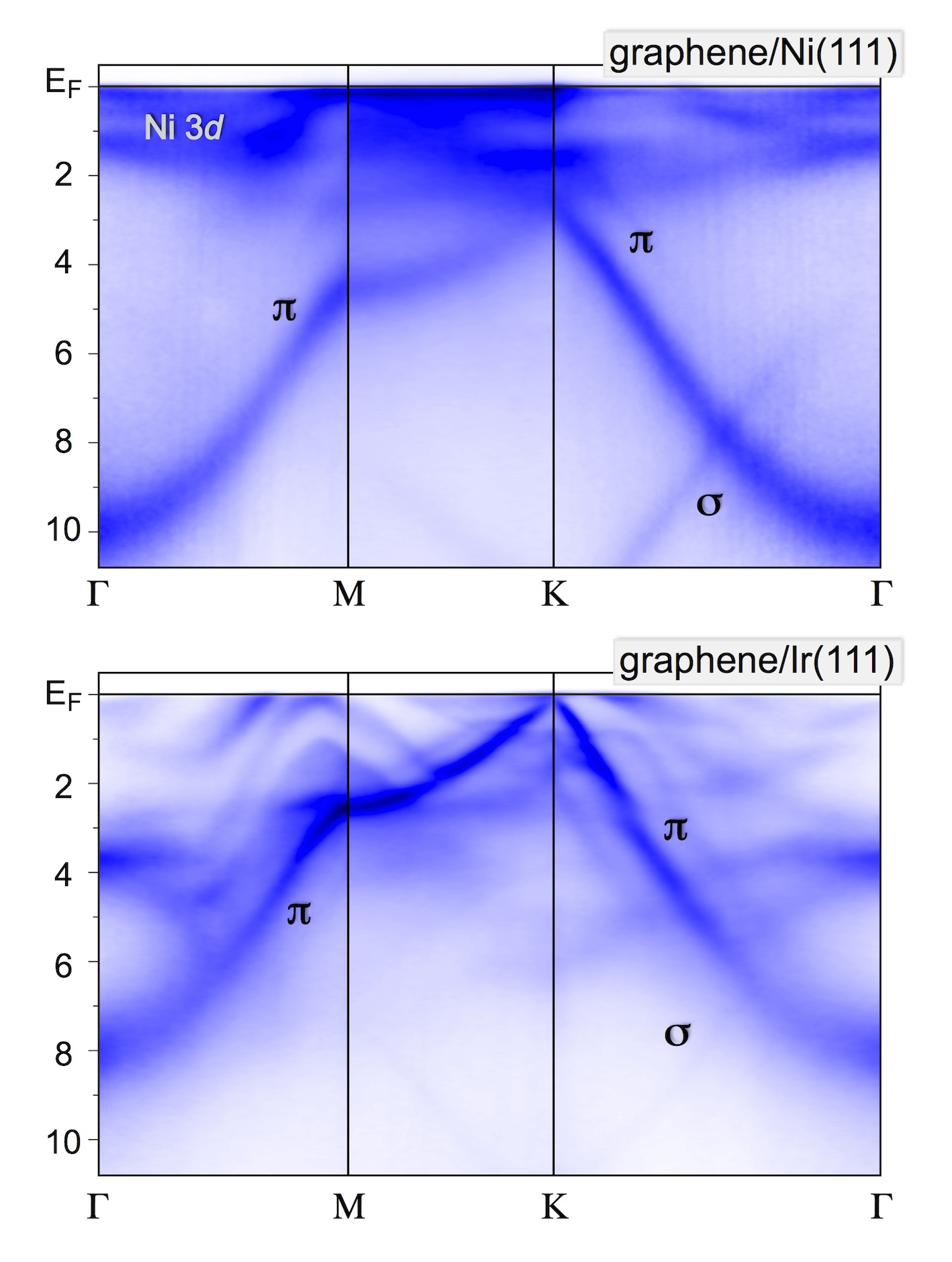}\\
\caption{\label{fig:arpes} ARPES intensity maps along the $\Gamma-M-K-\Gamma$ path in the Brillouin zone for graphene/Ni(111) (upper panel) and graphene/Ir(111) (lower panel).}
\end{figure}

The graphene/Ni(111) system is an intensively-studied representative example of the lattice-matched system. The STM and LEED experiments show the commensurate $(1\times1)$ structure of carbon atoms on Ni(111) (Fig.~\ref{fig:stm}, left)~\cite{Dedkov:2010a}. The lattice constant extracted from the experiment is $2.4\pm0.1$\,\AA, which agrees well with the in-plane lattice constant of graphite ($2.464$\,\AA). Earlier studies of this system considered the first intuitively prospective arrangement of carbon atoms, namely $ATOP$. This model was supported by available experimental studies of graphene/Ni(111) by means of surface extended energy-loss fine-structure spectroscopy~\cite{Rosei:1983}. These experiments also found the relatively large distance of $2.46$\,\AA\ between a graphene layer and Ni(111) without any intermixing of graphene and Ni valence band states, that was in contradiction with available photoemission experiments. However, the later studies performed by means of LEED $I(V)$-curve analysis found that $HCP$ ($top-fcc$) arrangement of carbon atoms above Ni(111) is more realistic with the mean distance between carbon and top Ni layers of $2.135$\,\AA~\cite{Gamo:1997}. This arrangement was later supported by many electronic structure calculations, ranging from LSDA to GGA and GGA-D and now this model is widely accepted for the arrangement of carbon atoms on the Ni(111) surface~\cite{Bertoni:2004,Weser:2011,Voloshina:2011NJP,Adamska:2012}. The similar conclusions were made for graphene on Co(0001)~\cite{Eom:2009} and Fe/Ni(111)~\cite{Sun:2010,Weser:2011}.

The relatively small distance between graphene and Ni(111) leads to the significant intermixing of the valence band states of both materials: the strong hybridization between graphene\,$\pi$ and Ni\,$3d$ states was observed in the experiment [Fig.~\ref{fig:arpes}, upper panel]~\cite{Dedkov:2010a,Dedkov:2008a,Dedkov:2001,Weser:2010,Weser:2011,Voloshina:2011NJP,Gruneis:2008,Rusz:2010} and confirmed by theoretical results~\cite{Bertoni:2004,Weser:2011,Voloshina:2011NJP,Karpan:2007,Karpan:2008}. Here the electronic structure of free-standing graphene is drastically modified: for the occupied valence band states the graphene-derived $\sigma$ and $\pi$ states are shifted to higher binding energies by 1\,eV and 2.4\,eV, respectively, due to the charge transfer from Ni states onto unoccupied states of graphene; the graphene Dirac cone is not preserved due to the strong intermixing of Ni and graphene valence band states and the large gap is opened around the $K$ point of the graphene-derived Brillouin zone. The corresponding changes are also observed for the unoccupied valence band states. Following the above description, presently, the graphene/Ni(111) is related to the system where graphene is ``strongly'' bonded to metallic substrate. The analogous description is also accepted for other lattice-matched systems, graphene/Co(0001) and graphene/Fe/Ni(111).

The lattice-matched graphene/Cu(111) system prepared via intercalation of thin Cu layer underneath graphene can be related to the ``weakly'' interacting system due to the completely filled $d$ shell. Here the recent theoretical and experimental works show the weak interaction of graphene\,$\pi$ and Cu valence band states~\cite{Dedkov:2001,Khomyakov:2009,Varykhalov:2010a}. Also the Dirac cone is preserved in the system. The similar effects were observed for the system where graphene was grown by molecular-beam epitaxy of carbon atoms on single crystalline Cu(111)~\cite{Walter:2011fj}. 

In the case of lattice-mismatched graphene-based systems two cases can be identified: (i) ``weakly'' bonded graphene to metal, graphene/Ir(111), graphene/Pt(111) and (ii) ``strongly'' bonded graphene to metal, graphene/Ru(0001), graphene/Rh(111). 

The graphene/Ir(111) is a widely studied example of a ``weakly'' bonded graphene on the lattice-mismatched metallic substrate. Its representative STM image is shown in Fig.~\ref{fig:stm} (right image) in the inverted contrast imaging when $ATOP$ positions are viewed as dark and $FCC$ and $HCP$ as a bright regions~\cite{Ndiaye:2006,Coraux:2008,Coraux:2009}. The conclusion regarding the ``weak'' type of interaction in the system under consideration was made on the basis of its almost intact electronic structure compared to free-standing graphene: (i) the binding energy of the graphene-derived $\pi$ states at the $\Gamma$ point is $8.15$\,eV, (ii) the $\pi$ band has a linear dispersion around $E_F$, (iii) there is a small $p$-doping of graphene in this system [Fig.~\ref{fig:arpes}, lower panel]~\cite{Pletikosic:2009,Starodub:2011a}. Additionally the moir\'e superstructure of graphene on Ir(111) results in modulation of the periodic atomic potential around carbon atoms followed by the appearing of the Dirac cone replicas and the opening of minigaps in the band structure of a graphene layer~\cite{Pletikosic:2009}. Experiments performed on this system by means of x-ray standing wave (XSW) technique yield a mean distance between graphene and Ir(111) equal to $3.38$\,\AA\ with a corrugation between $0.6$ and $1.0$\,\AA~\cite{Busse:2011}. From the theoretical point of view the LDA approach gives the surprisingly good result for the mean distance between graphene and Ir(111) of $3.42$\,\AA~\cite{Feibelman:2008,Feibelman:2009}, although one has to be aware that this  can be due to the error cancellation and this method has a tendency to overestimate the binding. The GGA calculations, on the contrary, yield a very small binding energy of only several meV per carbon atom with a relatively large mean distance between graphene and Ir(111)~\cite{Ndiaye:2006,Lacovig:2009}. The recent development of the methods which can take into account the non-local van der Waals interactions allowed to  describe the graphene/Ir(111) system correctly~\cite{Busse:2011,Voloshina:2012b}. The obtained mean distance and the corrugation of the graphene layer are in the reasonable agreement with experimental data. In this DFT-D description the graphene layer is bonded to Ir(111) via the van der Waals interaction with an antibonding average contribution from chemical interaction~\cite{Busse:2011}. These calculations indicate the charge accumulation at the graphene/Ir(111) interface with the small charge transfer of $\approx0.01$\,electrons/C from graphene to Ir resulting in a slight $p$-doping of graphene with the Dirac cone shifted by $0.2$\,eV above $E_F$, which is consistent with experimental value of $0.1$\,eV~\cite{Busse:2011,Pletikosic:2009}. The similar observations and conclusions are also valid for the ``weakly'' interacting graphene/Pt(111) system~\cite{Land:1992,Sasaki:2000,Ueta:2004,Sutter:2009a,Gao:2011,MartinezGalera:2011kw,Gyamfi:2012eg}. The later results are also consistent with the model calculations for the doping level taking into account the difference in the work function of graphene and the metal support~\cite{Giovannetti:2008,Khomyakov:2009}.

\begin{figure}
\centering
\includegraphics[width=0.8\textwidth]{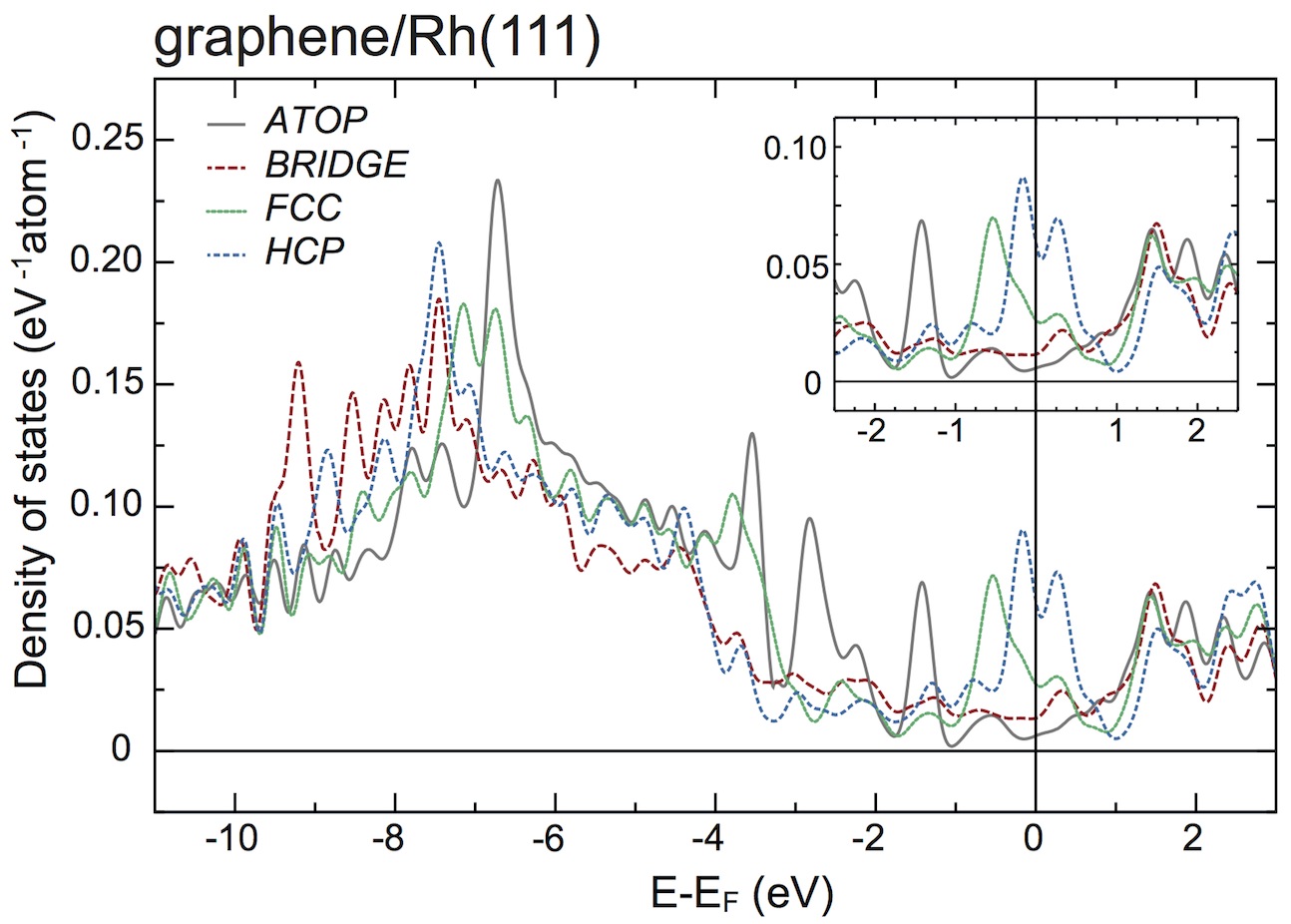}\\
\caption{\label{fig:grRh_dos} Carbon atom-projected total density of states ($\sigma$ and $\pi$) in the valence band for the different high-symmetry positions of the graphene/Rh(111) system. The inset shows the corresponding density of states for the $p_z$ character only.}
\end{figure}

Graphene on Rh(111) is chosen by us as a typical example of the ``strongly'' bonded lattice-mismatched system. Its STM and LEED images are shown in the middle part of Fig.~\ref{fig:stm}~\cite{Wang:2010ky,Sicot:2012,Voloshina:2012a}. The moir\'e structure is clearly visible in the experiment. The lattice constant of the moir\'e structure extracted from the experimental data is $\approx29$\,\AA\, corresponding to the arrangement of $(12\times12)$ graphene layer on $(11\times11)$ Rh(111). All high-symmetry adsorption sites of carbon atoms on Rh(111) are marked on STM image by the corresponding capital letters and these positions can be unambiguously assigned on the basis of comparison of calculated and experimental STM data~\cite{Iannuzzi:2011iq,Voloshina:2012a}. The measured corrugation of the graphene layer on Rh(111) depends on the STM imaging conditions and was found to vary in the range $0.5-1.5$\,\AA\ depending on imaging conditions~\cite{Wang:2010ky,Sicot:2012,Voloshina:2012a}. The DFT-D optimized structure of graphene/Rh(111) is presented in Fig.~\ref{fig:grRh_struct}(a) and the variation of the hight of the carbon atoms is shown in Fig.~\ref{fig:grRh_struct}(b). Carbon atoms in the $ATOP$ configuration define a high-lying region sitting at $d_0=3.15$\,\AA\ above Rh(111), and those in other configurations form a lower region. The buckling in the graphene overlayer is $1.07$\,\AA. Carbon atoms in the $BRIDGE$ configuration form the lowest topographic area ($d_0=2.08$\,\AA). The $HCP$ and $FCC$ regions are approximately  $0.4$\,\AA\ and  $0.8$\,\AA\ higher than the minima. Thus, the theoretically obtained value of corrugation is in very good agreement with those obtained in atomic force microscopy (AFM) measurements of this system of $1.1$\,\AA~\cite{Voloshina:2012a}. Authors of Ref.~\cite{Voloshina:2012a} have estimated the influence of dispersion forces on the obtained results: while qualitatively the observed picture remains the same, non inclusion of the van der Waals interactions (i.\,e. standard DFT-PBE treatment) yields larger corrugation ($\approx1.8$\,\AA) with a very similar low region ($d_0=2.10$\,\AA), but a high region at $d_0=3.90$\,\AA. This is due to the alternating ``weak'' and ``strong'' interactions of graphene with the Rh(111) surface. In the case of the ``strong'' interaction between graphene and metal, standard GGA-treatment gives a reasonable result, whereas for the areas of the ``weakly'' bonded graphene dispersion forces, neglected by the standard procedure, are important.

\begin{figure*}
\centering
\includegraphics[width=1\textwidth]{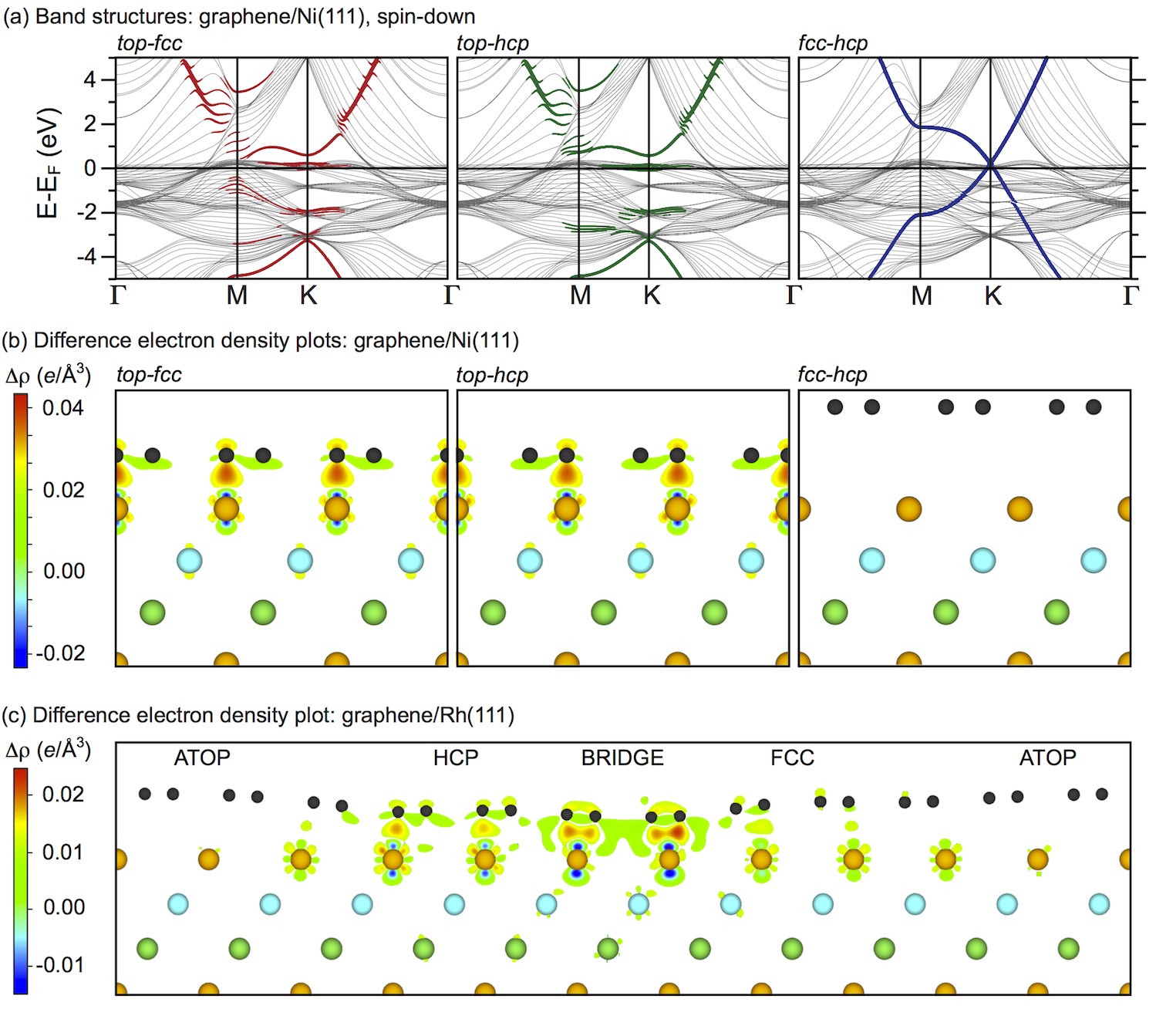}\\
\caption{\label{fig:grNi_vs_grRh} (a) Electronic band structure for different arrangements of a graphene layer on Ni(111). (b) and (c) show the difference electron density, $\Delta \rho(r) = \rho_\mathrm{gr/Rh}-\rho_\mathrm{Rh}-\rho_\mathrm{gr}$, plotted in units of $e/\AA^3$ calculated for different positions of graphene/Ni(111) and graphene/Rh(111). Red (blue) colors indicate regions where the electron density increases (decreases).}
\end{figure*}

The coexistence of the ``weakly'' and ``strongly'' bonded graphene on Rh(111) is reflected in the electronic structure. As can be seen from Fig.~\ref{fig:grRh_dos}, where projected DOSs on the representative C atoms are shown, the low-lying region of graphene/Rh(111) displays rather a strong interaction involving hybridization of graphene states with those of the metal, whereas for the high-lying region a free-standing graphene-like picture can be observed (although small doping is visible). The difference in DOS curves around the Fermi level is not so pronounced as expected due to the fact that for the $ATOP$ positions of a graphene layer it is close to the free-standing one with the low DOS at $E_F$ (the finite DOS is due to the residual interaction with the Rh substrate). In case of the $BRIDGE$ positions, the interaction is so strong that it might open a large gap in DOS for the graphene $\pi$-states. But the number of hybrid interface states can lead to the increasing of DOS in the vicinity of the Fermi level. This situation is analogous to the graphene/Ni or graphene/Co lattice-matched systems, where such situation is observed [Fig.~\ref{fig:grNi_vs_grRh} (a)]~\cite{Bertoni:2004, Weser:2011,Voloshina:2011NJP,Eom:2009}. The difference in the DOS peak positions for the graphene $\sigma$- and $\pi$-states reflects the strength of hybridization and the charge transfer in the system. Analogous to the above description, the strong interaction between graphene and Rh around the $BRIDGE$ positions can be explained by the strong hybridization between graphene $\pi$ and Rh $d$ states in the valence band. That can lead to the partial charge transfer from Rh to graphene. This effect is reflected in DOS as a shift of $\pi$-states of graphene from $\approx6.5$\,eV binding energy for $ATOP$ position to higher binding energies for $BRIDGE$ positions (several DOS peaks in the energy range $E-E_F=-7.5\ldots-9.5$\,eV). The peaked resonances at the Fermi level can be viewed as the signature of an active dangling bond. This suggests that the $HCP$ and $FCC$ area display the strongest chemical activity in the moir\'e overstructure and therefore  one might expect preferable formation of TM clusters above these regions~\cite{Wang:2011hh}.

\begin{figure*}
\centering
\includegraphics[width=1\textwidth]{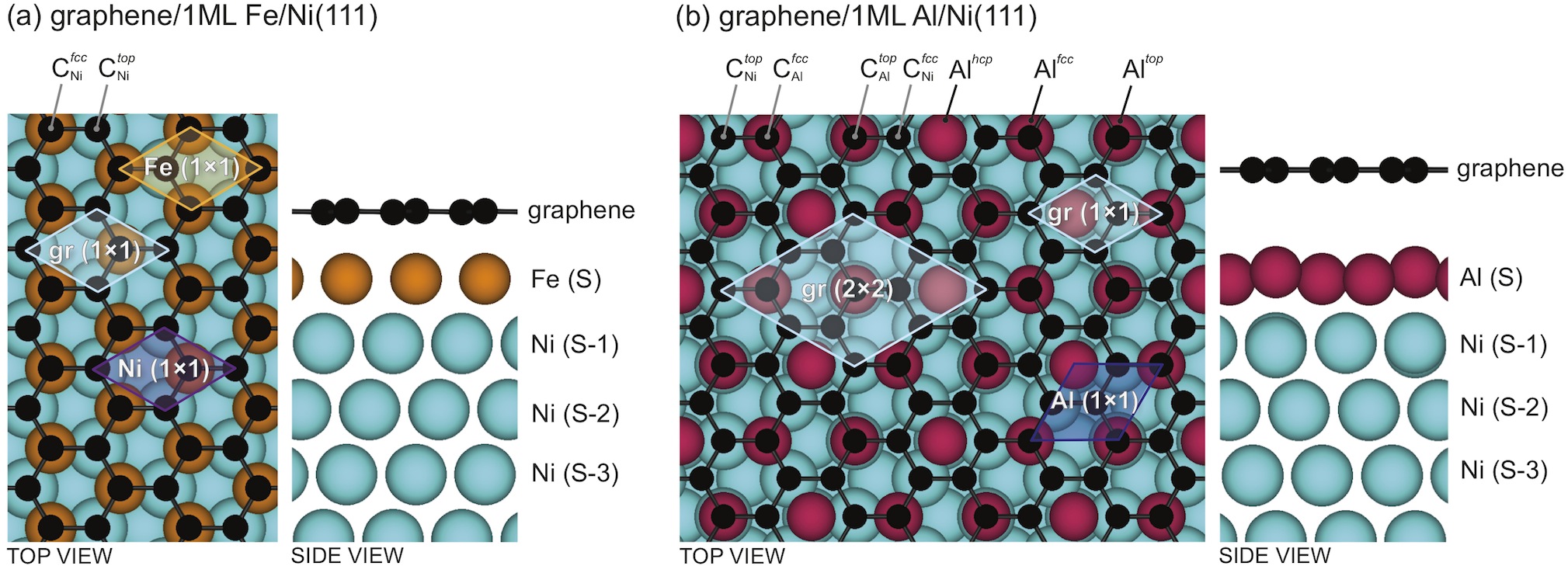}\\
\caption{\label{fig:inter-struct} Side and top view of the crystallographic structures of the (a) graphene/Fe(111) and (b) graphene/Al/Ni(111) intercalation-like systems. The corresponding units cells are marked in the figure.}
\end{figure*}

Now we will try to make a connection between electronic structure and geometry of lattice-matched and lattice-mismatched graphene/metal systems. For the gra\-phe\-ne/Ni(111) interface the electronic properties, strength of hybridization and level of doping of graphene depend on the arrangement of the system. For the $top-fcc$ ($HCP$) and $top-hcp$ ($FCC$) arrangements of carbon atoms on Ni(111) the ``strong'' interaction and significant hybridization of graphene\,$\pi$ and Ni\,$3d$ states are observed [Fig.~\ref{fig:grNi_vs_grRh}(a), left and middle plots]. Here, these effects lead to the opening of the energy gap between graphene\,$\pi$ and $\pi^*$ states with appearing of many, so called, interface states in this gap. For the $fcc-hcp$ ($ATOP$) arrangement of carbon on Ni(111) the Dirac cone is preserved and in this case the graphene layer is even slightly $p$-doped by 200\,meV above $E_F$ [Fig.~\ref{fig:grNi_vs_grRh}(a), right plot]. Surprisingly, this value of doping is close to the one for the graphene layer on Ir(111)~\cite{Pletikosic:2009} (see also discussion above in the text). We would like to emphasize, that for all possible arrangements of graphene on Ni(111) the symmetry for two carbon sublattices in the graphene unit cell is broken, meaning that the opening of the band gap in the graphene spectra is expected in all cases. However, as is evident from Fig.~\ref{fig:grNi_vs_grRh} the breaking of the symmetry is not the only prerequisite for the opening of the gap in the graphene layer. The strength of the additional modulation potential (in this case, the strength of site-specific hybridization between valence band states) is important for the gap opening. This effect of hybridization, space and energy overlapping, of the graphene\,$\pi$ and Ni\,$3d$ valence band states is clearly demonstrated in Fig.~\ref{fig:grNi_vs_grRh}(a,b) where for the $top-fcc$ ($HCP$) and $top-hcp$ ($FCC$) configurations the hybridization states appear between a graphene layer and the Ni(111) surface. For the $fcc-hcp$ ($ATOP$) case the distance between graphen and Ni (111) is large preventing space overlapping of orbitals leading to the preservation of the Dirac cone. As an outlook, the two factors, the real space and the energy overlapping of the hybridized states defines the appearance and the value of the energy gap in graphene.

For the graphene/Rh(111) interface the situation is more complex. Here, all possible high-symmetry positions (see discussion above) are possible. In this case one can expect the different local hybridization strength as well as local electron doping of the graphene layer. This point is supported by the electronic structure calculations for this system [Figs.~\ref{fig:grRh_dos} and ~\ref{fig:grNi_vs_grRh}(c)]. The calculated local DOS for the graphene/Rh(111) system is discussed earlier and it gives the idea about the local hybridization strength of the valence band states of graphene and Rh and the local doping. Considering the local charge distribution map [Fig.~\ref{fig:grNi_vs_grRh}(c)], the similarity between systems, graphene/Rh(111) and different arrangements in graphene/Ni(111), is clearly visible. The $ATOP$, $HCP$, and $FCC$ stackings of graphene on Rh(111) and Ni(111) have practically the same distribution of the electron density: (i) formation of interface states and (ii) charge transfer from metallic substrate to graphene for $HCP$ and $FCC$, (iii) strong charge localization at the interface between graphene and metal, (iv) absence of any visible interaction for the $ATOP$ positions. Therefore, in the case of graphene/Rh(111) one may expect that the local electronic structure for these places will be similar to those of respective arrangements of graphene/Ni(111) [Fig.~\ref{fig:grNi_vs_grRh}(a)]. The most locally strongly interacting place for the graphene/Rh(111) system is the $BRIDGE$ position, which is, however, is not stable for graphene/Ni(111)~\cite{Adamska:2012}. For this place the charge transfer and the interface charge localization is even stronger compared to the $HCP$ and $FCC$ places [Fig.~\ref{fig:grNi_vs_grRh}(c)]. The nearly similar situation is realized for the graphene/Ru(0001) system, where the ``strong'' interaction between graphene and Ru is observed as well~\cite{Brugger:2009}. The opposite case is graphene/Ir(111) where interaction is ``weak'' and almost no hybridization between graphene and Ir valence band states as well as small charge transfer are observed~\cite{Busse:2011}. However, we would like to emphasize that for all lattice-mismatched graphene/metal systems all above-listed high-symmetry positions for carbon atoms are realized.    

Summarizing all these facts one can expect that for the lattice-mismatched ``strongly-interacting'' systems in the electronic structure extracted from, for example, photoemission experiments all possible replicas of all crystallographic stackings have to be detected: two or more parabolas for graphene\,$\pi$ or $\sigma$ states. However, available experimental data demonstrate the absence of such structures and single band always exist either for $\pi$ or for $\sigma$ graphene-derived states~\cite{Brugger:2009,Dedkov:2012a,Brugger:2009,Pletikosic:2009,Rusponi:2010}. These observations can be explained by the metallic nature of graphene in these systems. The high mobility of electrons and that during description of the electronic structure such lattice-mismatched systems have to be considered as a whole object without separation on ``weakly'' and ``strongly'' bonded regions lead to the fact that a graphene layer becomes fully doped via the charge transfer from the metallic substrate. 

The broken symmetry in the carbon lattice as well as locally distributed strong hybridization between graphene and substrate valence band states lead to the appearing of the relatively large gap around the $K$ point of the graphene-derived Brillouin zone. This picture seems to be valid for ``strongly'' bonded graphene on Rh(111) and Ru(0001), where in $FCC$ and $HCP$ regions the symmetry for two carbon sublattices is locally broken and strong hybridization of the graphene\,$\pi$ and $d$ states of a substrate exists. For the ``weakly'' bonded graphene on Ir(111) the local interaction between graphene and Ir substrate is small everywhere in the moir\'e lattice and no strong hybridization is observed, meaning no gap around the $K$ point. The slight doping of graphene in this system is defined only by the charge transfer from graphene on Ir(111). The perturbation of the graphene-lattice potential by the weak periodic potential of the moir\'e lattice leads only to the appearing of weak replicas in the photoemission spectra with small gaps where they intersect with the main photoemission branches from graphene.

\subsection{Graphene-metal-based intercalation-like systems}

\begin{figure}[h]
\centering
\includegraphics[width=0.6\textwidth]{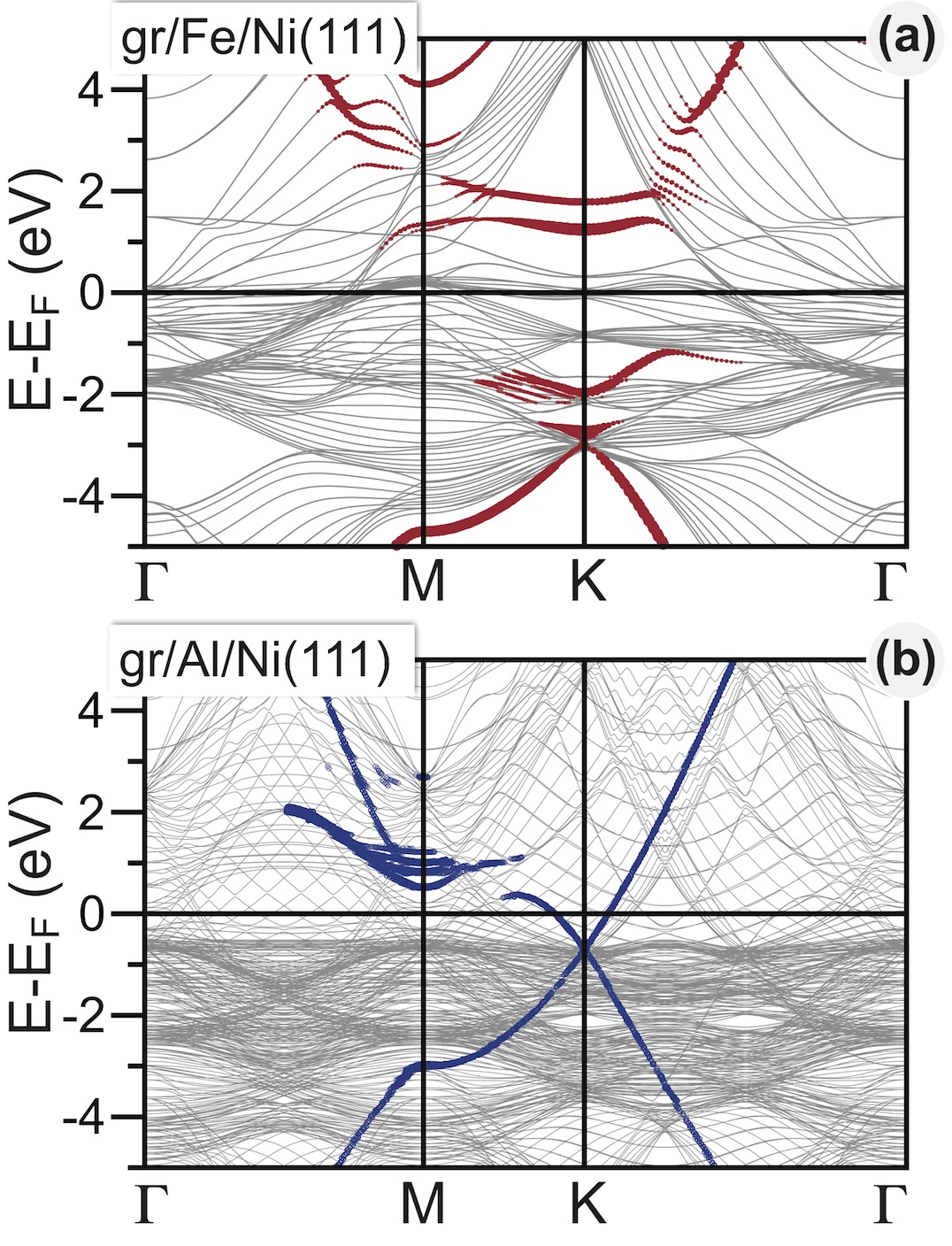}\\
\caption{\label{fig:inter-bands} Minority-spin electronic band structures calculated for the (a) graphene/Fe/Ni(111) and (b) graphene/Al/Ni(111) intercalation-like systems. The graphene\,$\pi$ character in both plots is emphasized by thicker lines.}
\end{figure}

\begin{figure*}[t]
\centering
\includegraphics[width=1\textwidth]{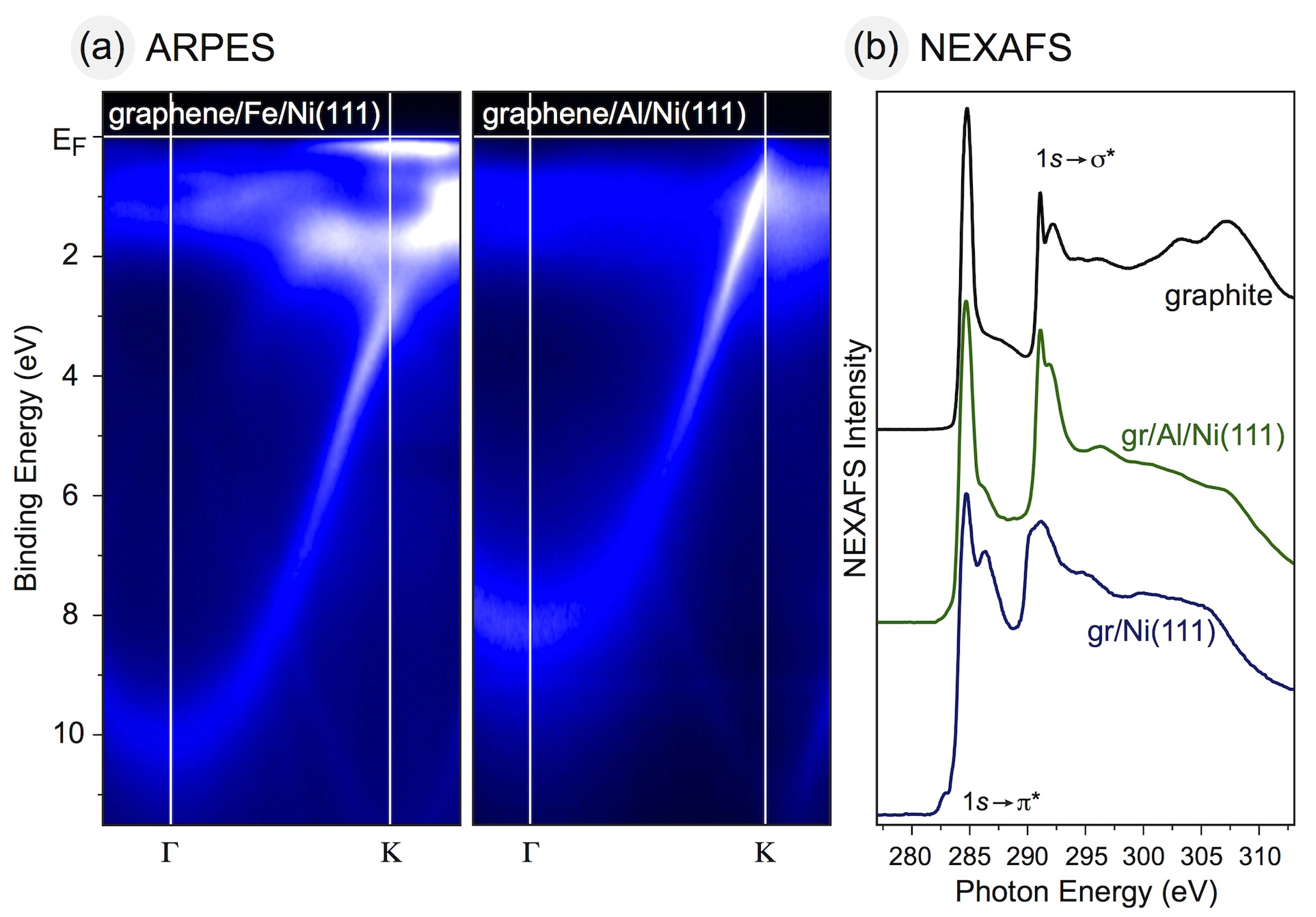}\\
\caption{\label{fig:arpesNEXAFS} (a) ARPES photoemission intensity maps for graphene/Fe/Ni(111) and graphene/Al/Ni(111) presented for the $\Gamma-K$ direction of the graphene Brillouin zone. (b) C $K$ edge NEXAFS spectra of graphene/Ni(111), graphene/Al/Ni(111), and a graphite single crystal.}
\end{figure*}

The properties of epitaxial graphene/substrate interfaces can be efficiently controlled by inserting other materials between graphene and the original support, a process referred to as intercalation. In case of the graphene/metal system, different examples were investigated, which can be found in the recent review papers~\cite{Wintterlin:2009,Batzill:2012,Dedkov:2012book}. Considering the graphene/Ni(111) interface as a reference system, we would like to present here a few illustrative examples and the first one is dealing with the graphene/Fe/Ni(111) intercalation-like system. The lattice structure, which was revealed by comparative analysis of the experimental (LEED) and theoretical (DFT/PBE) results~\cite{Weser:2011}, is shown in Fig.~\ref{fig:inter-struct} (left): Fe atoms are placed in the $fcc$ hollow sites following thereby the Ni(111) structure and C atoms are staying in the \textit{top-fcc} arrangement with respect to the Ni(111) substrate. The interaction between graphene and metal is stronger here than in the graphene/Ni(111) case due to the less filled $d$ shell of Fe compared to Ni. Intercalation of iron underneath of the graphene layer changes drastically the magnetic response from graphene that is explained by the formation of the highly spin-polarized $3d_{z^2}$ quantum-well state in thin iron layer, keeping the electronic structure below the Fermi energy almost intact [compare Figs.~\ref{fig:grNi_vs_grRh} (a) and \ref{fig:inter-bands} (a) for theory (DFT/PBE), and Figs.~\ref{fig:arpes} and \ref{fig:arpesNEXAFS}, for experiment (ARPES)].

Thus, in the previous example, the unique conical electronic structure of graphene was completely destroyed. One can try to restore it via decoupling the graphene layer from metal, Ni(111), by intercalation of Cu or Au, for example~\cite{Shikin:2000,Dedkov:2001,Varykhalov:2008,Varykhalov:2010a}. Indeed, one may observe the desired picture, although the Dirac point is usually shifted  up- or downwards with respect to the Fermi level, resulting in $p$- or $n$-doped graphene, respectively. Additionally, $\pi$ states of graphene are hybridized with $d$-states of the noble metal~\cite{Varykhalov:2008,Varykhalov:2010a}. The latter effect can be avoided, if necessary, by replacing Cu-Au with $sp$-metal, like Al.

The crystallographic arrangement of atoms in the graphene/Al/Ni(111) system is presented in Fig.~\ref{fig:inter-struct}. Contrary to the graphene/Fe/Ni(111) trilayer, both occupied and unoccupied states are modified by means of Al-intercalation~\cite{Voloshina:2011NJP,Generalov:2012gi,Addou:2012}. As one can see from Figs.~\ref{fig:arpes} and \ref{fig:inter-bands}, in this case all electronic bands of graphene are shifted  to lower binding energies, compared to graphene/Ni(111). Furthermore, the electronic structure of the graphene layer as well as the Dirac cone in the vicinity of $E_F$ are fully restored (there is a small electron doping of graphene leading to a shift of the Dirac point below $E_F$ by ca.\,$0.64$\,eV). The decoupling process in this case is also easily visible when comparing the near edge x-ray absorption fine structure (NEXAFS) spectra obtained for graphite, graphene/Ni(111) and graphene/Al/Ni(111) presented in Fig.~\ref{fig:arpesNEXAFS}. In the case of our reference system, graphene/Ni(111), in the region of the $1s \rightarrow\pi^*$ transition, this spectrum has a double-peak structure compared to the one of graphite that is explained by the transitions of C $1s$ core electron into two unoccupied states (interface states), which are the result of hybridization of the graphene and Ni valence band states~\cite{Weser:2010,Rusz:2010}. Additionally, one observes a reduction in the energy separation between the $\pi^*$ and $\sigma^*$ features compared to that in the spectra of graphite (a result of the lateral bond softening within the adsorbed graphene monolayer). Intercalation of thin Al layer underneath a graphene layer on Ni(111) leads to drastic changes in the C $K$ NEXAFS spectrum: The shape of the spectrum, positions of main spectroscopic features as well as the energy separation between $\pi^*$ and $\sigma^*$ features become similar to those in the spectrum of pure graphite. These facts immediately indicate that the graphene layer is decoupled from the substrate by intercalated Al. An interesting feature of the system under consideration is that owing to the symmetry breaking in the graphene lattice there is no energy gap for the $\pi$ states around the $K$ point of the Brillouin zone of graphene and the both theory and experiment agree in this statement. At the same time, appearance of such a gap was recently demonstrated by ARPES for the graphene layer on Cu(111), Ag(111) and Au(111)~\cite{Varykhalov:2010a,Enderlein:2010,Walter:2011fj}. Note, however, that these experiments are not supported by band-structure calculations where a very small~\cite{Kang:2010} or no energy gap was observed for these surfaces~\cite{Khomyakov:2009}.

Thus, by intercalation of different metals underneath graphene on Ni(111) one can (i) shift the graphene unoccupied states away from $E_F$, when employing Co or Fe~\cite{Dedkov:2011wa,Weser:2011} or (ii) decouple the graphene states from the Ni(111) substrate, when employing noble metals or Al~\cite{Shikin:2000,Dedkov:2001,Varykhalov:2010a,Enderlein:2010,Voloshina:2011NJP}. In this way, in the latter case hybridization between graphene $\pi$ and metal valence band states in the vicinity of $E_F$ is completely avoided.

\section{Discussion}

Coming to the end of the present manuscript we will rise and discuss two main problems in the studying of graphene layers on metallic substrates. Going through the presently available literature which is mainly summarized in several review papers~\cite{Wintterlin:2009,Batzill:2012,Dedkov:2012book} one can come to the conclusion that this field of graphene surface science research gives probably the only way for the preparation of high-quality graphene layers which can be either used in future devices in the contact with the supporting metal or can be transferred on the insulating or semiconducting support without losing the fundamental properties of graphene~\cite{Yu:2008,Kim:2009a,Bae:2010,Li:2009}. Here the controllable growth as well as high quality of obtained graphene layers open the door for the wide implementation of these technology in industry~\cite{Yang:2012,Lee:2012a}. Along with this rapid progress in the application of graphene, the nature of interaction of graphene with metals is far from the full understanding.

As it was pointed earlier, initially the application of standard DFT methods (LSDA, GGA-PBE) gave in most cases the correct description of the geometrical arrangement of a graphene layer on metallic substrates. However, the bonding strength was either overestimated (LSDA) or significantly underestimated up to non-bonding situations (GGA-PBE). When corrected for the missing dispersion interaction, the accuracy of DFT is currently becoming acceptable and one can predict the bonding energies as well as the distances between graphene and metallic substrates (for lattice-matched and lattice-mismatched systems). 

The separation of the graphene/metal systems on two subclasses, lattice-matched and lattice-mismatched, does not give an answer on the different strength of interaction. In the subclass of lattice-mismatched systems both situations, ``strongly'' and ``weakly'' interacting graphene, are observed [graphene/Rh(111) vs. graphene/Ir(111)]. The first attempt to describe the difference in the bonding between graphene and metals was made in the framework of the so-called $d$ band model~\cite{Hammer:2000wd}, which predicts a stronger binding with decreasing occupation of the $d$ band giving a stronger binding when one goes in one group from the $5d$ to the $3d$ metals. According to this model the changes in the properties of the graphene/metal interface have to be more gradual as it is observed for the $3d$ and $4d$ metals. But the drastic changes in the bonding between graphene and Rh and Ir or graphene and Pd and Pt are observed. For example, the DFT studies by means of GGA-PBE predict that graphene is bonded to Ru, but repulsive interaction is predicted for adsorption of graphene on PtRu$_2$, Pt$_2$Ru, and Pt surfaces as well as on Ir~\cite{Okamoto:2005,Brako:2010}. This effect probably could be understood in terms of orbital overlap. In going from $4d$ to $5d$ elements, relativistic effects become significant, resulting in contracting the $6s$ orbitals and expanding the $5d$ orbitals. Probably, the overlap of the graphene\,$\pi$ states with more diffuse $5d$ orbitals is worse than with more localized $4d$ orbitals. Later on, the effect of bonding of graphene on Pt(111) and Ir(111) was checked by the inclusion of van der Waals forces in calculations~\cite{Busse:2011,Ugeda:2011ci}. However, even these state-of-the-art calculations do not give a clear intuitive answer on the bonding mechanism between graphene and metallic surfaces. 

In this situation, the so-called \textit{incremental scheme}~\cite{Stoll:1992}, which belongs to the group of local correlation methods [see e.g.\,Refs.~\cite{Pulay:1983,Hampel:1996}], may help considerably~\cite{Paulus:2006,Voloshina:2011ck,Voloshina:2007,Voloshina:2012}. An asset of this approach is the possibility to analyze the individual correlation contributions to the binding energy, i.\,e. one can separate interactions between, e.\,g., $s$- or $d$-orbitals of metal and those of graphene, that will give a better insight in the nature of bonding between graphene and substrate.

Another interesting point concerning the metal/graphene interface is the existence of the energy gap at the $K$ point in the graphene\,$\pi$-electron spectrum. The appearance of such gap is closely related to the violation of the symmetry in the graphene lattice. For the ``strongly'' interacting graphene/metal interfaces [graphene/Ni(111), graphene/Co(0001), graphene/Ru(0001)] the existence of such gap is unequivocally described by the theory, both GGA-PBE and GGA-PBE-D, that also supported by the experimental observations. Here, as it was described earlier, the gap opening is due to the broken symmetry for two carbon sublattices along with the strong hybridization between graphene\,$\pi$ states and $d$-states of the substrate.

For the ``weakly'' interacting graphene/metal interfaces the situation is not so obvious. In the beginning we would like to recall the situation on the existence of such gap for the $n$-doped graphene on SiC(0001)~\cite{Rotenberg:2008,Zhou:2008}. It was demonstrated that the appearance of such gap in the spectrum of graphene, which was observed in Ref.~\cite{Zhou:2008} and assigned to the violation of the sublattice symmetry, can be assigned to the islanded graphene sample which could be obtained during underanneling of SiC substrate~\cite{Rotenberg:2008}. Moreover, the additional indication of the broken symmetry in the graphene layer is the observation of the intensity from the graphene\,$\pi$ band in the second Brillouin zone along $\Gamma-K$ direction. As was shown by Shirley \textit{et al.}~\cite{Shirley:1995}, if the potential for both carbon sublattices is identical, then due to the interference between emission from A and B sites of the graphene unit cell the intensity in the second Brillouin zone vanishes. In opposite case the final photoemission has to be detected.

In the simple ``weakly'' interacting graphene/Ir(111) and graphene/Pt(111) the energy gap around $K$ was not detected in experiment or predicted by theory~\cite{Pletikosic:2009,Sutter:2009a}. It was possible to open a gap only via strong hydrogenation of a graphene layer on Ir(111) when local transformation from $sp^2$ to $sp^3$ hybridization of carbon occurs~\cite{Balog:2010} or via heavy co-doping with Ir and Na~\cite{Papagno:2011hl}. In the later case this rehybridization of carbon atoms due to adsorption of Ir in $HCP$ sites of moir\'e structure can be locally enhanced by strong charge transfer in the presence of adsorbed Na.

In case of the graphene/metal intercalation-like systems the earlier experiments demonstrate either existence of the gap, like for graphene/Na,K,Cs/Ni(111)~\cite{Nagashima:1994}, or its absence, like for graphene/Cu,Au/Ni(111)~\cite{Shikin:2000,Dedkov:2001}. However the correctness of these old results can not be taken as a reference due to the experimental limitations in former time (limited energy and angular resolutions), because the fact of detection of such gap is very crucial to the method used. For example, later experiments performed with a modern display-type photoelectron detectors on graphene/K/Ni(111) system demonstrate that intercalated K leads to heavy $n$-doping of graphene, decoupling of graphene from substrate, lack of any hybridization of graphene\,$\pi$ and substrate valence band states and the absence of any gap around the $K$ point for the graphene\,$\pi$ states~\cite{Gruneis:2008}.

The recent ARPES experiments performed on the graphene/Cu/Ni(111), graphene/Ag/Ni(111), and graphene/Au/Ni(111) intercalation-like systems~\cite{Varykhalov:2010a} indicate an energy gap around $K$ of $180$\,meV, $320$\,meV, and $0$\,meV, respectively. This is related by the authors of this work to the broken symmetry in the system, after noble metal intercalation, connected with the doping level of graphene in the obtained system. However, the available band-structure calculations on the level of LSDA do not reproduce these observations~\cite{Giovannetti:2008,Khomyakov:2009}. First of all this discrepancy between theory and experiment can be attributed to the non-correct description of the graphene/metal interface in the framework of LSDA which always overestimates the binding. But in this case the theoretically predicted splitting has to be even larger. Also, the recent work on the graphene/Al/Ni(111) system do not show any gap for $\pi$ states around the $K$ point in experimental as well as in theoretical data~\cite{Voloshina:2011NJP}, although the doping level of graphene is quite significant with the Dirac point localized by $\approx0.64$\,eV below $E_F$. The existence of the energy gap for $\pi$ states was also demonstrated for graphene/Au/Ru(0001) system~\cite{Enderlein:2010} that was supported by the observation of the non-vanishing intensity of the $\pi$ band in the second Brillouin zone (see discussion above). Nevertheless, this gap was detected after depositing on the initial graphene/Au/Ru(0001) system (initially is $p$-doped) K atoms, that was attributed in Ref.~\cite{Varykhalov:2010a} to the effect of potassium doping of the system. Here, we would like to mention also several recent works where no or extremely small gap (if any) was observed for the ``weakly'' bonded graphene/metal systems~\cite{Gierz:2008,Gierz:2010,Gao:2010a,Gao:2011,Meng:2012ee}. Results of these works point out that for the formation of the energy gap for graphene\,$\pi$ states the effect of broken symmetry in the graphene unit cell between two carbon sublattices has to be accompanied with the effect of strong modulation of the carbon periodic potential due to the hybridization of $\pi$ states with the valence band states of substrate. In the ideal case the space, energy, and wave vector conservations during hybridization have to be fulfilled.

\section{Conclusions}

The aim of the present perspectives article is to give a short overview of the main problems in the field of surface physics and chemstry of the graphene/metal interface. The main questions which are still open, concerning this system, are the origin of drastic variations of interactions in these systems (two subclasses) as well as an origin of modification of electronic structure of graphene in the vicinity of the $K$ point. We have tried to organise these questions in a more ordered way and prepare roads for the solution of these problems. Several examples of the graphene/metal interfaces together with related intercalation-like systems were considered here from both experimental and theoretical points of view, that allows us to shed more light on the problem of the interaction of graphene with the substrates and modification of its electronic structure. At the same time, the recent progress in experimental and theoretical investigations of the graphene/metal interfaces opens a wide door in the application of these systems in future electron and spin transport devices. 

\section*{Acknowledgement}

Y.\,D. would like to thank Prof. V. K. Adamchuk, Prof. A. M. Shikin, Prof. K. Horn, Dr. M. Fonin, M. Weser for the intense collaboration in various joint graphene-related projects. E.\,V. would like to thank Prof. B. Paulus for giving the support during this research. The authors acknowledge Dr. N. Gaston for the careful reading the manuscript. The computing facilities (ZEDAT) of the Freie Universit\"at Berlin and the High Performance Computing Network of Northern Germany (HLRN) are acknowledged for computer time.

\end{document}